\documentclass[pra,10pt,twocolumn,superscriptaddress,showpacs]{revtex4-1}
\usepackage{amsmath}
\usepackage{latexsym}
\usepackage{amssymb}
\usepackage{graphicx}
\usepackage[colorlinks=true, citecolor=blue, urlcolor=blue]{hyperref}
\usepackage{float}
\usepackage{amsfonts}
\usepackage{textcomp}

\begin{document}
\title{Entanglement renders free riding redundant in the thermodynamic limit}
\author{Shubhayan Sarkar}
\author{Colin Benjamin}
\email{colin.nano@gmail.com}
\affiliation{National Institute of Science Education and Research, HBNI, Jatni- 752050, India}

\begin{abstract}
The free rider problem is one of the most well studied problems in economics. The solution proposed mainly is punitive in order to deter people from free riding. In this work we introduce quantum strategies and also study the problem by using the mathematical structure of 1D Ising model in the infinite player limit. We observe that for maximum entanglement the quantum strategy is always the equilibrium solution, i.e., the strategy chosen by majority of the players independent of the payoffs, solving the free rider problem.
\end{abstract}
\maketitle
\section{Introduction}
 In economics, the public goods game is typically used to investigate cooperative behavior. In the public goods game, a resource is produced which is perfectly shareable, and once produced can be utilized by all in a community. In a community, the number of members can be quite large, thus can be taken as infinite or the thermodynamic limit. Since the resource is a public good, people have less of an incentive to provide than to free ride, known as the well-known free rider problem. This problem, in the thermodynamic limit, is usually tackled numerically \cite{19}. In a recent work \cite{10}, the classical two-player public goods was extended to the thermodynamic limit and solved analytically. In this paper, the similarity between 1D Ising model and game theory in the thermodynamic limit, as established in Refs. \cite{10,3,qnash} is used to derive the equilibrium strategy for the quantum public goods game in the thermodynamic limit.

We use the same technique as used in Refs.~\cite{3,10,qnash}, where sites are replaced by players and the spin states at each site are analogous to the player's strategic choices. Similar to Ising model, magnetization in the thermodynamic limit of a game is defined as the difference in fraction of players choosing strategy, say $s_1$ over $s_2$. We first review the analogies between two spin Ising Hamiltonian and a two-player game and then extend it to the thermodynamic limit. The public goods game is then quantized using Eisert's scheme\cite{2} and the payoffs corresponding to the strategies be it quantum or classical are calculated. The quantum public goods game is then extended to the thermodynamic limit by considering the strategies quantum versus classical, i.e., quantum versus provide or quantum versus free ride. We see that for the quantum public goods game, quantum and provide are equally possible. Further, when quantum goes against free ride, an interesting feature emerges, for the maximally  entangled case, the majority always chooses the quantum strategy and do not free ride, independent of the payoffs. We also check the behavior of quantum Nash equilibrium as the amount of entanglement in a site is changed.

This paper is organized as follows- in section IA, we explain the 1D Ising model and the connection between a two spin Ising Hamiltonian and the payoffs of a classical two-player two-strategy game as was done in Ref.~\cite{1}. The thermodynamic limit of the game is also defined as was done in Refs.~\cite{10,qnash}. Next, the public goods game is quantized using Eisert's scheme\cite{2} and the Nash equilibrium strategy is calculated for the quantum public goods game. We plot the difference in the fraction of players choosing a particular strategy versus the entanglement $\gamma$ at a particular site and observe phase transitions. Interestingly, we find that in the quantum public goods games when the entanglement between the players in a particular site is maximal, the overwhelming majority of players choose the quantum strategy independent of the payoffs. 
\subsection{Mapping classical game payoffs to Ising parameters}\label{sec2}
In statistical physics, 1D Ising model\cite{6} represents a system of spins arranged on a line which can be in either of two states $-1$ (down-spin:$\downarrow$) or $+1$ (up-spin:$\uparrow$) and only interact with their nearest neighbors. The 1D Ising Hamiltonian is given as-
\begin{equation}\label{eq10}
H=-J\sum^N_{i=1}\sigma_i\sigma_{i+1}-h\sum^N_{i=1}\sigma_i,
\end{equation}
 $J$ denotes the spin-spin coupling, $h$ represents the applied magnetic field and the spins are represented by $\sigma$'s while $i$ denotes the site index. Using Hamiltonian Eq.~(\ref{eq10}), one can derive the magnetzation which is- 
\begin{equation}\label{eq8}
m=\frac{\sinh(\beta h)}{\sqrt{\sinh^2(\beta h)+e^{-4\beta J}}}.
\end{equation}
An 1D Ising system with two spins can be mapped to a two-player game as has been shown in Refs.~\cite{1,3}. The payoffs for two-player game can be written in matrix form as-
\begin{equation}\label{eq9}
U=\left(\begin{array}{c|cc} & s_1 & s_2 \\\hline s_1 & a,a' & b,b' \\  s_2 & c,c' & d,d'\end{array}\right),
\end{equation} 
wherein $U(s_i,s_j)$ denotes the payoff function and $a, b, c, d$ denote payoffs for row player while $a', b', c', d'$ denote the payoffs for column player. $s_1$ and $s_2$ represent the strategic choices made by the two-players. Following on from Refs. \cite{3,10,1}, the Ising game matrix is defined as the energy states of the two site Ising chain-
\begin{equation}\label{ising-game}
\left(\begin{array}{c|cc}  & s_2=+1 & s_2=-1  \\\hline s_1=+1 & J+h & -J+h \\s_1=-1 & -J-h & J-h\end{array}\right).
\end{equation} 
with $s_1,\ s_2$ being the spin states at the two sites. The players are analogous to the sites whereas their strategic choices are analogous to spins in Ising model. One can map Eq.~(4) to Eq.~(3) and then get $J, h$ in terms of $a, b, c, d$, thus going to the thermodynamic limit of the game, since the magnetzation for the infinite site 1D Ising model is derived in terms of the Ising parameters $J, h$. Thus for the game matrix in the  infinite site (or thermodynamic) limit, the spin-spin coupling $J$ and external magnetic field $h$ are given as-  
\begin{eqnarray}\label{eq5}
J=\frac{a-c+d-b}{4},\ h=\frac{a-c+b-d}{4}.
\end{eqnarray} 
in terms of payoffs of a general two-player game. The magnetization, i.e., the difference in average number of players choosing strategy $s_1$ against the average number of players choosing $s_2$, from Eq.~(\ref{eq8}) in the thermodynamic limit with $J$ and $h$, as in Eq.~(\ref{eq5}) can then be written as-
\begin{equation}\label{eq13}
m=\frac{\sinh(\beta \frac{a-c+b-d}{4})}{\sqrt{\sinh^2(\beta \frac{a-c+b-d}{4})+e^{-\beta (a-c+d-b)}}}.
\end{equation} 
Decreasing $\beta$ (implies increasing temperature) in Ising model increases the disorderness in orientation of the spins. Thus, decreasing $\beta$ in the  magnetization of game theory, Eq.~(\ref{eq8}) increases randomness in a player's strategic choices. For connecting quantum public goods game with 1D Ising model we proceed as follows. We first quantize the classical public goods game and then remodel the 1D Ising spin chain incorporating entanglement at each site. 
\section{Public Goods game}
The public goods game or the "free rider problem" is a social dilemma game \cite{8,7}. In the two-player  public goods game, the "public good" is produced by either player alone by paying the entire cost of the good or both of the players pay for half of the good. The payoffs for the cooperators (provider) and defectors (free rider) \cite{3,7} are given as-
\begin{eqnarray}
P_D=kn_cc/N,\ \ P_C=P_D-c,
\end{eqnarray}
where $c$ is the cost of the good, $k$ denotes the multiplication factor of the "public good", $N$ denotes the number of players in the group (in the two-player public goods game, $N=2$), and $n_c$ denotes the number of cooperators in the group. The punishment $p$ is introduced such that whenever a player defects or free rides he has an additional negative payoff given by $-p$. Thus, the payoff matrix can be written as is-
\begin{equation}\label{eq27}
U=\left(\begin{array}{c|cc} & provide & free\ ride \\\hline provide & 2r,2r & r-\frac{c}{2},r+\frac{c}{2}-p \\  free\ ride & r+\frac{c}{2}-p,r-\frac{c}{2} & -p,-p\end{array}\right),
\end{equation}
where $r,c,p>0$. As we can see from the payoff matrix Eq.~(\ref{eq27}), when $r>c/2-p$, then cooperation or provide is the Nash equilibrium but when $r<c/2-p$ then defection or free riding is the Nash equilibrium. When $p=0$, the above game Eq.~(\ref{eq27}) reduces to public goods game without punishment. 
Note: Although in this work we only consider the thermodynamic or infinite player limit our results are equally valid for a mena field model with considerably lesser number of players. Since our approach is based on the 1D Ising model, we would look at the scenario in the same context to figure out the $N$ where our results would also be valid in the mean field case. From ID Ising model, the partition function is given by
\begin{eqnarray}
Z&=&\lambda_+^N+\lambda_-^N= \lambda_{+}^N (1+\alpha)\nonumber
\end{eqnarray}
where $\lambda_{\pm}=\cosh(\beta h)\pm \sqrt{\sinh^{2}(\beta h) + e^{-4\beta J}}$ are the eigenvalues of the transfer matrix (see Ref.~[7] for details regarding the derivation of the partition function for Ising model) and $\alpha=(\frac{\lambda_{-}}{\lambda_{+}})^N$. Since, the relation $\lambda_{-}<\lambda_{+}$ is always true, thus for large enough $N$, $Z=\lambda_+^N$. Now, Taking $\log$ on both sides,
\begin{eqnarray}
N \log(\frac{\lambda_{-}}{\lambda_{+}})=\log(\alpha)\implies N=-\log(\alpha)/\log(\frac{\lambda_{+}}{\lambda_{-}})\nonumber
\end{eqnarray}
Now, for our approach to be valid $\alpha \rightarrow 0$. Thus, the analytic approach is valid for any $N$ such that $\alpha$ could be neglected which depends on $\beta h$ and $\beta J$. For example if $ J=1.5/\beta, h=0.3/\beta$, then say for $\alpha=10^{-6}$, we get $N\sim 20$ which means the mean field limit is reached for no. of sites $N>20$. Now since we have a one to one correspondence between the Ising model and our public goods game, substituting the Ising parameters $J$ and $h$ in terms of the game payoffs as derived in Eq. (5) above and a very small value for $\alpha=10^{-6}$, then
\begin{widetext}
\begin{equation}
N=-\log(\alpha)/\log(\frac{\cosh(\beta \frac{a-c+b-d}{4})+ \sqrt{\sinh^2(\beta \frac{a-c+b-d}{4})+e^{-\beta (a-c+d-b)}}}{\cosh(\beta \frac{a-c+b-d}{4})- \sqrt{\sinh^2(\beta \frac{a-c+b-d}{4})+e^{-\beta (a-c+d-b)}}})
\end{equation}
\end{widetext}
 We have the payoffs $a,b,c,d$ for the public goods game given by $a=2*reward/\beta, b=(reward-cost/2)/\beta, c=(reward-punishment+cost/2)/\beta, d=-punishment/\beta$. Taking $reward=1, cost=4$ and $punishment=0$, we have: $N\sim 15$. This gives the mean field  limit for the game can be reached with around $15$ sites\cite{9}.\\
\\ 
\subsection{Mapping quantum public goods game payoffs to Ising parameters}
To quantize the classical two-player public goods game, we use the scheme proposed by Eisert, et. al., see~\cite{2} in the context of Prisoner's dilemma. The two-players are represented by two qubits while the strategies they adopt are represented by the state of the qubits. The cooperation or provide strategy is represented as $|0\rangle$ while defect or free ride strategy is represented as $|1\rangle$. Entanglement in the game is introduced by an operator $L$ which entangles the state of the qubits. $L$ is given by-
\begin{equation}
L=\left(\begin{array}{cccc} \cos(\gamma/2)& 0 & 0 & i \sin(\gamma/2)  \\  0& \cos(\gamma/2) & -i \sin(\gamma/2) & 0 \\  0 & -i \sin(\gamma/2)  & \cos(\gamma/2)& 0 \\ i \sin(\gamma/2) & 0 & 0 & \cos(\gamma/2) \end{array}\right)\nonumber
\end{equation}
with $\gamma$ representing the amount of entanglement between qubits. $\gamma=0$ implies no entanglement while $\gamma=\pi/2$ defines a maximally entangled state at each site. Thus, the initial state after action of $L$ on $|00\rangle$ is given as- 
$|\psi_w\rangle= \cos(\gamma/2)|00\rangle+i\ \sin(\gamma/2)|11\rangle$, where the subscript $w$ indicates site index. Players choose a particular strategy by applying the unitary operator $O(\theta,\phi)$ on the initial state $|\psi_w\rangle$, wherein
\begin{equation}\label{unitary}
O(\theta,\phi)=\left(\begin{array}{cc} e^{i\phi}\cos(\theta/2) & \sin(\theta/2) \\ -\sin(\theta/2) & e^{-i\phi}\cos(\theta/2) \end{array}\right).
\end{equation} The final state which results from the action of  disentanglement operator $L^{\dagger}$ and the unitaries $O(\theta_{i},\phi_{i}), i=A,B$ representing the strategies of players/qubits $A$ and $B$ is- 
\begin{equation}
|\chi_w\rangle=L^{\dagger}O(\theta_{A},\phi_{A})\otimes O(\theta_{B},\phi_{B})L|00\rangle
\end{equation}
The classical public goods game payoffs are given in Eq.~(\ref{eq27}). The payoffs for players $A$ and $B$ are then calculated using the classical public goods game payoffs as-
\begin{eqnarray}
\$_A=2r|\langle00|\chi_w\rangle|^2+(r+\frac{c}{2}-p)|\langle10|\chi_w\rangle|^2\nonumber\\+(r-\frac{c}{2})|\langle01|\chi_w\rangle|^2-p|\langle11|\chi_w\rangle|^2, \mbox{ for player A,}\nonumber\\
\$_B=2r|\langle00|\chi_w\rangle|^2+(r-\frac{c}{2})|\langle10|\chi_w\rangle|^2\nonumber\\+(r+\frac{c}{2}-p)|\langle01|\chi_w\rangle|^2-p|\langle11|\chi_w\rangle|^2, \mbox{ for player B}\nonumber.
\end{eqnarray}  
The provide operation is defined by identity matrix as in Eq.~\ref{unitary}, while free ride is defined by $X$. The quantization procedure  allows us to introduce the quantum strategy given by $Q=iZ=O(0,\pi/2)$. The new payoff matrix including the quantum strategy is-
\begin{small}
\begin{equation}\label{eq17}
U=\left(\begin{array}{c|ccc} & provide & free\  ride & Q\\\hline provide & 2r,2r & r-\frac{c}{2},r+\frac{c}{2}-p  & \alpha 1,\alpha 1\\  free\ ride & r+\frac{c}{2}-p,r-\frac{c}{2} & -p,-p & \alpha 2,\alpha 3 \\Q & \alpha 1,\alpha 1 & \alpha 3,\alpha 2 &  2r,2r\end{array}\right)
\end{equation}
\end{small}
wherein $\alpha 1=2r\cos^2(\gamma)-p\sin^2(\gamma)$, $\alpha 2=r-\frac{c}{2}\cos (2\gamma)-p\sin^2(\gamma)$ and $\alpha 3=r+\frac{c}{2}\cos(2\gamma)-p\cos^2(\gamma)$. For maximally entangled state, $\gamma=\frac{\pi}{2}$, the payoff matrix taking reward:$r=1,$ cost $c=4$ and punishment $p=0$, i.e., quantum public goods game without punishment is-
\begin{small}
\begin{equation}\label{value-nop}
U=\left(\begin{array}{c|ccc} & provide & free\  ride & Q\\\hline provide & 2,2 & -1,3  & 0,0\\  free\ ride & 3,-1 & 0,0 & 3,-1 \\Q & 0,0 & -1,3 &  2,2\end{array}\right)
\end{equation}
\end{small}
The Nash equilibrium for the quantum public goods game without punishment, as seen from Eq.~(\ref{value-nop}) for maximal entanglement is $(free ride,free ride)$. On the other hand, the payoff matrix for quantum public goods game with punishment, taking reward $r=1$, cost $c=4$  and punishment $p=1$, for maximal entanglement is-
\begin{small}
\begin{equation}\label{value-p}
U=\left(\begin{array}{c|ccc} & provide & free\  ride & Q\\\hline provide & 2,2 & -1,2  & -1,-1\\  free\ ride & 2,-1 & -1,-1 & 2,-1 \\Q & -1,-1 & -1,2 &  2,2\end{array}\right)
\end{equation}
\end{small}
The Nash equilibrium for the quantum public goods game with punishment, as seen from Eq.~(\ref{value-p}) for maximal entanglement is $(Q,Q)$.
Remarkably, in the next section we will see that regardless of punishment in the thermodynamic limit, quantum is the equilibrium solution.
\section{Quantum public goods game in the thermodynamic limit}
\begin{figure}[h!]
\begin{center}
\includegraphics[width=\linewidth]{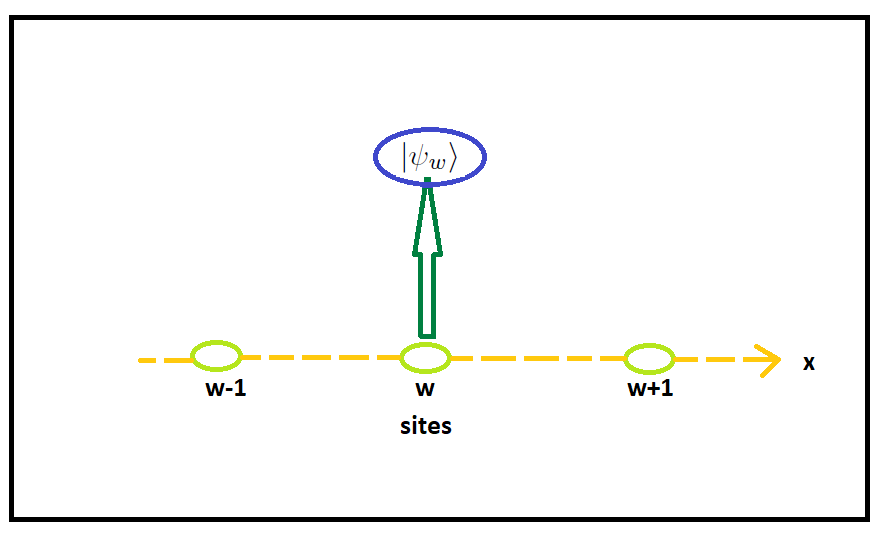}
  \caption{The quantum public goods game is played at each site in the 1D Ising model. At each of the $w$ sites a two-player quantum public goods game is played on an entangled state $|\psi_{w}\rangle$. The yellow line represents the classical coupling $J$.   The other Ising parameter $h$ which is the external magnetic field, plays the role of aligning various sites towards a particular strategy.}
  \label{fig8:}
  \end{center}
\end{figure}

We extend the two-player Quantum public goods game, discussed in the previous section, to the  thermodynamic limit via a mapping to the Ising model as discussed in section IA. We use a similar approach as in Ref.~\cite{8} wherein each Ising site consists of an entangled pair and different sites interact via classical couplings $J$. In Fig.~\ref{fig8:}, a schematic diagram is shown where at each site a two-player quantum public goods game is played. In classical Ising model, the applied magnetic field $h$ aligns the spins onto a specific orientation, similarly in quantum public goods game in the thermodynamic limit, $h$ plays the role of an external parameter which creates a bias in player's choices for the infinite-site(thermodynamic limit) quantum game. Since, in Ising model the sites can have only two states, we have each player in a game with access to either a classical or quantum strategy, see Eq.~(\ref{unitary}). We solve for the thermodynamic limit of the game and find the magnetization Eq.~(\ref{eq8}) when classical and quantum strategies are adopted at each site.
As was argued in Ref.~\cite{8}, and since we want to investigate the classical strategies versus quantum, we break Eq.~(\ref{eq17}) into two separate $2\times2$ games and first analyze quantum versus provide strategy and then the quantum versus free ride strategy.
\subsubsection{Quantum versus Provide}
As discussed in section~\ref{sec2} we calculate the magnetization for the case when players can choose either  quantum ($Q=iZ$) or provide ($C=I$) strategies. Thus, the payoffs from Eq.~(\ref{eq17}) for the row player (the column player's payoffs can be deduced from Eq.~(\ref{eq17}) is- 
\begin{small}
\begin{equation}
U=\left(\begin{array}{c|cc} & provide & Q \\\hline provide & 2r & 2r \cos^2(\gamma)-p\sin^2(\gamma) \\  Q & 2r \cos^2(\gamma)-p\sin^2(\gamma) & 2r\end{array}\right).
\end{equation}
\end{small} 
From the payoff matrix, it can be inferred that there are two Nash equilibrium- both choose provide or both choose quantum. To make the connection with Ising game matrix, see Eq.~(\ref{ising-game}), we transform the matrix as given above using the method explained in section 1A, thus
$J=\frac{(2r+p) \sin^2(\gamma)}{2}$ and $h=0$. The magnetization then is-
\begin{equation}
m=\frac{\sinh(\beta h)}{\sqrt{\sinh^2(\beta h)+e^{-4\beta J}}}=0.
\end{equation} 
As we can see  for the thermodynamic limit of quantum public goods game when the players have access to quantum or provide strategies, the magnetization becomes $0$. This means that there are equal proportions of players choosing quantum and provide strategy and thus can be concluded that players equally choose between quantum and provide strategy. This is similar to the two-player version.    
\subsubsection{Quantum versus Free ride}
As discussed in section \ref{sec2} we calculate the magnetization for the case when players make either  quantum ($Q=iZ$) or free ride ($freeride=X$) as their strategic choices. The payoff matrix in this case is given by-
\begin{small}
\begin{equation}
U=\left(\begin{array}{c|cc} &Q&free ride\\\hline Q &2r&\frac{2r-{c}\cos (2\gamma)-2p\sin^2(\gamma)}{2}\\free ride&\frac{2r+{c}\cos(2\gamma)-2p\cos^2(\gamma)}{2}&-p\end{array}\right)
\end{equation}
\end{small}
The Nash equilibrium in this case as can be seen from Eq.~(16) depends on $\gamma$. For $r>\frac{c}{2}\cos (2\gamma)-p\sin^2(\gamma)$, the Nash equilibrium is both players choosing quantum. However, when $r<\frac{c}{2}\cos (2\gamma)-p\sin^2(\gamma)$, the Nash equilibrium is both players choosing to free ride or defect. As derived in section 1A, the Ising parameters further are $J=0$ and
$h=\frac{r-\frac{c}{2}\cos(2\gamma)+p\cos^2(\gamma)}{2}$. The magnetization is then-
\begin{eqnarray}\label{eq34}
m&=&\frac{\sinh(\beta h)}{\sqrt{\sinh^2(\beta h)+e^{-4\beta J}}},\nonumber\\
\mbox {or, }m&=&\tanh(\beta(\frac{r-\frac{c}{2}\cos(2\gamma)+p\cos^2(\gamma)}{2})).
\end{eqnarray} 
\begin{figure}[t!]
\begin{center}
\includegraphics[width=.9\linewidth]{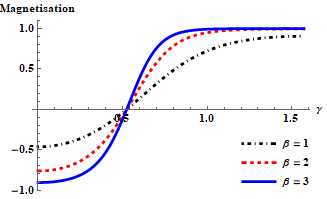}
  \caption{magnetzation versus $\gamma$ for $r=1$, $c=4$ and $p=0$ for different $\beta's$.  For the  maximally entangled case i.e. $\gamma=\pi/2$, the magnetzation always turns out to be positive independent of $r,\ c,\ p$ and $\beta$.}
  \label{fig9:}
  \end{center}
\end{figure}
\begin{figure}[t!]
\begin{center}
\includegraphics[width=.9\linewidth]{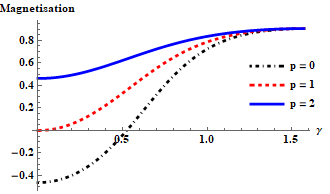}
  \caption{magnetzation versus $\gamma$ for $r=1$, $c=4$ and $\beta=1$ for different $p$.}
  \label{fig10:}
  \end{center}
\end{figure}
The quantum strategy reduces to the provide or cooperation strategy when $\gamma=0$ for public goods game with punishment. Thus, for $\gamma=0$, magnetzation from Eq.~(\ref{eq34}) is-
\begin{eqnarray}
m=\tanh(\beta(\frac{r+p}{2}-\frac{c}{4})).\nonumber
\end{eqnarray} 
which is the same equation for the classical public goods game with punishment in the thermodynamic limit as was derived in Ref.~\cite{10}. A phase transition for the quantum public goods game with punishment would occur when-
\begin{eqnarray}\label{eq38}
r-\frac{c}{2}\cos(2\gamma)+p\cos^2(\gamma)=0,\mbox{i.e.,} \cos^2(\gamma)=\frac{2r+c}{2(c-p)}.\nonumber\\
\end{eqnarray}
From Eq.~(\ref{eq38}) the phase transition occurs at $\gamma=\cos^{-1}\sqrt{(1+2)/4}=\pi/6$ (for $r=1$, $c=4$ and $p=0$) as can be clearly seen from Fig \ref{fig9:}. From Fig \ref{fig10:}, we see that as the punishment $p$ increases for all values of $\gamma$, the majority of population would always choose quantum. Further, it should be noted from Eq. (\ref{eq34}) that  the magnetization always comes out to be non-negative when $\gamma=\pi/2$ independent of payoffs in Eq. (\ref{eq27}) as for public goods game $r>0$. Thus, independent of the reward, cost or  punishment, majority of the players always opt for the Quantum strategy. From Fig \ref{fig9:}, it can be inferred that when $\beta$ increases and the magnetzation is positive, the fraction of players choosing quantum strategy also increases. However, when magnetzation is negative, the fraction of defectors increases as $\beta$ increases. When $\beta \rightarrow 0$, the players tend to become unbiased and equally choose both strategies. 
\section{Conclusions}
The aim of this work was to extend the two-player public goods game to the thermodynamic limit and derive the strategy adopted by the majority of players. We observe that in the thermodynamic limit, the provide and quantum strategy are always equi-probable and the players are unbiased towards either of the choices. However, for the case when players are allowed to choose either free ride or quantum strategy, a phase transition can be seen when the entanglement present at each site increases. Further, for the maximally entangled case, the majority of players always opt for quantum strategy and do not free ride given any cost or punishment. Thus we can with confidence conclude that when the players are allowed to opt for quantum strategy, free riding in a population decreases thereby, solving the free-riding problem in public goods game. Our work to derive analytically in the thermodynamic limit a measure for the fraction of providers against the fraction of free riders can also be extended when there are loners too as in Ref.\cite{21} by invoking a mapping to a spin-1 Ising model will be attempted soon.

\end{document}